\documentstyle{menu95}
\begin{document}
\title{
NON--PERTURBATIVE STRUCTURE OF THE NUCLEON
}
\author{
      Ulf-G. Mei{\ss}ner \\
{\it Institut f\"ur Theoretische Kernphysik, Universit\"at Bonn,
Nussallee 14-16} \\
{\it D-53115 Bonn, Germany} \\
Preprint TK 95 20 \\
}
\maketitle

\begin{abstract}
Chiral perturbation theory is the effective field theory of the
standard model. In this talk, I discuss some applications of this
framework to the pion--nucleon system. These are chiral corrections
to the S--wave pion--nucleon scattering lengths, the reaction $\pi N
\to \pi \pi N$ at threshold and low--energy theorems in $\pi^0$
photoproduction.
\end{abstract}

\section{Effective Field Theory of QCD}

\noindent In the sector of the three light quarks ($u,d,s$), one can write
the QCD Lagrangian as
\begin{equation}
{\cal L}_{\rm QCD} = {\cal L}_{\rm QCD}^0 - \bar{q}\, {\cal M} \, q
\, \, , \label{lqcd}
\end{equation}
with $q^T = (u,d,s)$ and ${\cal M} = {\rm diag}(m_u , m_d , m_s)$ the
current quark mass matrix. The current quark masses are believed to be
small compared to the typical hadronic scale, $\Lambda_\chi \simeq 1$~GeV.
${\cal L}_{\rm QCD}^0$ admits a global chiral symmetry, i.e. one can
independently rotate the left-- and right--handed components of the
quark fields. This symmetry is spontaneoulsy broken down to its
vectorial subgroup, SU(3)$_{L+R}$, with the appeareance of eight
massless Goldstone bosons. The explicit chiral symmetry breaking due
to the quark mass term gives these particles, identified with the
pions, kaons and eta, a small mass. The consequences of the
spontaneous and the explicit chiral symmetry breaking can be
calculated by means of an effective field theory (EFT), called
chiral perturbation theory.\cite{wein,gl84} ${\cal L}_{\rm QCD}$ is
mapped onto an effective Lagrangian with hadronic degrees of freedom,
\begin{equation}
{\cal L}_{\rm QCD} = {\cal L}_{\rm eff}[U, \partial U , \ldots,  {\cal
  M} , N]
\, \, , \label{leff}
\end{equation}
where the matrix--valued field $U(x)$ parametrizes the Goldstones,
${\cal M}$ keeps track of the explicit symmetry violation and $N$
denotes matter fields (like e.g. the nucleon). While the latter are
not directly related to the symmetry breakdown, their interactions are
severely constrained by the non--linearly realized
chiral symmetry and one can thus incorporate
them unambiguously.\cite{cwz} ${\cal L}_{\rm eff}$ admits an energy
expansion,
\begin{equation}
{\cal L}_{\rm eff} = {\cal L}_{\pi \pi}^{(2)} + {\cal L}_{\pi \pi}^{(4)}
+ {\cal L}_{\pi N}^{(1)}+ {\cal L}_{\pi N}^{(2)}+ {\cal L}_{\pi N}^{(3)}
+ {\cal L}_{\pi N}^{(4)} + \ldots\, \, , \label{leffex}
\end{equation}
where the superscript $(i)$ refers to the number of derivatives and or
meson mass insertions. I restrict myself to the two--flavor
sector. The first two terms in Eq.\ref{leffex} comprise the meson
sector \cite{gl84} whereas the next four are relevant for processes
involving one single nucleon. The ellipsis stands for terms with more
nucleon fields and/or more derivatives. The various terms contributing
to a certain process are organized by their {\it chiral} dimension $D$
(which differs in general from the physical dimension) as follows:\cite{cnpp}
\begin{equation}
D = 2L + 1 \sum_d (d-2) N_d^{\pi \pi} + \sum_d (d-1) N_d^{\pi N} \, \,
,
\label{chidim}
\end{equation}
with $L$ the number of (pion) loops and $d$ the vertex dimension
(derivatives or factors of the pion mass). Lorentz invariance and
chiral symmetry demand that $d \ge 2$ ($\ge 1$) for mesonic
(pion--nucleon) interactions. So to lowest order, one has to deal with
tree diagrams ($L=0$) which is equivalent to the time--honored current
algebra (CA). However, we are now in the position of {\it systematically}
calculating the corrections to the CA results. It is also important to
point out  that ${\cal L}_{\pi \pi}^{(4)}$ and
${\cal L}_{\pi N}^{(2,3,4)}$ contain parameters not fixed by symmetry,
the so--called low--energy constants (LECs). These have to be
determined from data or can be estimated from resonance exchange.\cite{reso}
The whole machinery is well documented, see e.g. Ref.\cite{ulfrev}.

\section{Structure of the Nucleon}
\noindent Here, I wish to list some processes
which have been studied in detail
to give a flavor about where the CHPT machinery does apply in the
(single) nucleon sector. References can be traced back
from.\cite{bkmrev} It is worth to stress that tests of chiral dynamics
heavily rely on very {\it precise} {\it data} at {\it low} energies.
Fortunately, over the last
few years, such accurate data (for very different processes) have
become available and much more are coming.
\begin{itemize}

\item $\pi N \to \pi N$: Of particular interest are the S--wave
  scattering lengths and the so--called pion--nucleon
  $\sigma$--term (strangeness in the nucleon). In section 3,
  I will consider the chiral corrections
  to the threshold $\pi N$ amplitudes.

\item $\pi  N \to \pi \pi N$: This reaction has attracted particular
  interest since it supposedly allows to pin down the S-wave $\pi \pi$
  scattering lengths. The corresponding accurate
  CHPT predictions\cite{gl84} are one of the premier tests of chiral
  (Goldstone boson) dynamics. These issues are discussed in section 4.

\item $\gamma N \to \gamma N$: Low energy Compton scattering has been
   investigated in detail experimentally as well as theoretically over
   the last years. The empirical facts concerning the nucleons'
   electromagnetic polarizabilities find a natural explanation in
   CHPT. Furthermore, predictions for the spin--dependent amplitude
   have been made (spin--polarizability, slope of the generalized DHG
   sum rule). These predictions will be tested at BNL and
   CEBAF. Extensions to the three flavor sector have also been
   performed  and measurements using the hyperon beams at CERN and
   FNAL are eagerly awaited for.

\item $\gamma N \to \pi^0 N$: Here, it was shown that an existing
  low--energy theorem for the S--wave multipole $E_{0+}$ was incorrect
  and that the chiral expansion for the
  electric dipole amplitude is slowly converging. New data from MAMI
  and SAL (for $\gamma p \to \pi^0 p$)
  seem to indicate a smaller value (in magnitude) of $E_{0+}$
  in agreement with CHPT predictions. Also, novel P--wave LETs have
  been given and agree with indirect determinations. An accepted
  experiment at MAMI involving polarization will give a direct test.
  Also, the reaction $\gamma n \to \pi^0 n$ should be measured. A
  brief discussion is given in Section 5.

\item $\gamma^\star N \to \pi^a N$: Charged pion electroproduction is
  of particular interest since it allows to determine the axial form
  factor of the nucleon at small momentum transfer. A venerable LET
  due to Nambu et al. was modified and previously existing discrepancies
  to determinations of $G_A (t)$ from
  neutrino--nucleon reactions could be
  explained. Furthermore, the new NIKHEF and MAMI data on neutral pion
  electroproduction show some puzzling features which have yet to
  be understood.

\item $\mu p \to n \bar{\nu}_\mu$: Ordinary muon capture at rest allows to
  measure the induced pseudoscalar coupling constant $g_P$. In CHPT, a
  very accurate prediction can be made, $g_P = 8.44 \pm
  0.23$. Presently available determinations are not yet accurate
  enough to disentangle this from the simple pion pole (CA)
  prediction.

\end{itemize}
To end this short survey, I would like to stress again that all these
processes are to be considered in the threshold region, i.e. at small
energy and momentum transfer. Only there the CHPT machinery applies.

\section{The Isovector Pion--Nucleon S--Wave Scattering Length }

\noindent One of the most splendid successes of current algebra
was Weinberg's prediction for the S--wave pion--nucleon scattering
lengths,\cite{wein2}
\begin{equation}
a^+_{\rm CA} = 0 \, \, , \quad a^-_{\rm CA} = \frac{M_\pi}{8\pi
  F_\pi^2} \frac{1}{1+ M_\pi/m_p } = 0.079 \, M_\pi^{-1} \, \, ,
\label{CA}
\end{equation}
with $M_\pi = 139.57$ MeV the charged pion mass, $m_p = 938.27$ MeV the
proton mass, $F_\pi = 92.5$ MeV the pion decay constant and the
superscripts $+/-$ refer to the isoscalar and isovector $\pi N$
amplitude, respectively.
The Karlsruhe--Helsinki phase shift analysis of $\pi N$ scattering
\cite{karl} leads to
$a^- = 0.092 \pm 0.002 \, M_\pi^{-1}$ and $a^+= -0.008 \pm 0.004 \,
M_\pi^{-1}$, impressively close to the CA prediction, Eq.\ref{CA}.
However, over the last few years there has been some controversy about
the low--energy $\pi N$ data which has not yet been
settled. Consequently, the uncertainties in $a^{\pm}$ are presumably
larger and even the sign of $a^+$ could be positive. A more direct way
to get a handle at these zero momentum (i.e. threshold) quantities is the
measurement of the strong interaction shift ($\epsilon_{1S}$) and the
decay width ($\Gamma_{1S}$) in pionic atoms.
The PSI-ETH group has
recently presented first results of their impressive measurements in
pionic deuterium  and pionic hydrogen.\cite{pisd}
\cite{sigg} The consequent analysis of the data
leads to \cite{sigg}
$a^- = 0.096 \pm 0.007 \, M_\pi^{-1}$ and $a^+= -0.0077 \pm
0.0071 \, M_\pi^{-1}$. If one combines the pionic hydrogen shift measurement
with the one from the pionic
deuterium, one has
$a^- = 0.086 \pm 0.002 \, M_\pi^{-1}$ and $a^+= 0.002 \pm 0.001 \,
 M_\pi^{-1}$. The
largest uncertainty comes from the width measurement of pionic
hydrogen. Both determinations are consistent within one standard
deviation. We conclude that $a^-$ is larger than the CA value and that
$a^+$ is consistent with zero.

Within CHPT, the chiral corrections to Eq.\ref{CA} have been
calculated in Ref.\cite{bkma} There it was shown that the isoscalar
scattering length  is very sensitive to some LECs which are not known
to such an accuracy. In contrast, to order $M_\pi^3$ the only sizeable
corrections to $a^-$ come from the one loop diagrams, the counter term
contribution is small (as estimated from $\Delta (1232)$ and $N^\star
(1440)$ exchange). In Ref.\cite{bkmb} it was furthermore shown that
the one loop graphs with exactly one insertion from ${\cal L}_{\pi
  N}^{(2)}$ sum up to zero. Contact terms from ${\cal L}_{\pi
  N}^{(4)}$ can not contribute to $T^-$ due to crossing. Estimating
the uncertainties conservatively, one therefore arrives at a band
for $a^-$,
\begin{equation}
0.088 \, M_{\pi^+}^{-1} \le a^- \le 0.096 \,  M_{\pi^+}^{-1} \, \, ,
\label{amc}
\end{equation}
which is consistent with the various empirical values discussed
before and $10 \ldots 20\%$ larger than the CA prediction.\cite{wein2}
As already stressed in Ref.\cite{bkma}, it is the chiral loop
correction at order $M_\pi^3$ which closes the gap between the lowest
order (CA) prediction and the empirical value. An indication of the
size of the next corrections can be obtained by writing the one--loop
result as $a^- = a^-_{\rm CA} (1 + \delta_1) \simeq a^-_{\rm CA}
 \exp(\delta_1)$. The next correction follows to be $\delta_1^2 / 2$ which
 is of the order of $1 \ldots 2\%$ of $a^-_{\rm CA}$.

\section{{\boldmath $\pi N \to \pi \pi N$} at Threshold}

\noindent The reaction $\pi N \to \pi \pi N$
is of particular interest since it
contains, besides many other contributions, the four--pion
vertex. This offers the possibility to extract the S--wave $\pi \pi$
scattering lengths which are of fundamental importance to our
understanding of the chiral QCD dynamics. Over the last years, many
accurate threshold data have been compiled, but their theoretical
interpretation rested on the ancient Olsson--Turner model (which is
the same as tree level CHPT when one sets the
pre--QCD chiral symmetry breaking parameter $\xi =0$).
In Ref.\cite{bkmppnl} the first corrections to the threshold amplitudes
$D_1$ and $D_2$ were calculated and some novel LETs were formulated.
$D_{1,2}$ are related to the more commonly used ${\cal A}_{2I_{\pi
    N},I_{\pi \pi}}$ via
\begin{equation}
{\cal A}_{32} = \sqrt{10} \, D_1 \, \, , \quad
{\cal A}_{10} = -2 \, D_1 \, - 3 \, D_2 \quad ,
\label{curla}
\end{equation}
with $I_{\pi N}$  the total isospin of the initial pion--nucleon
system and  $I_{\pi \pi}$ the isospin of the final two--pion system.
These LETs show the expected pattern of deviation from the empirical
values, namely small and sizeable for ${\cal A}_{32}$ and ${\cal
  A}_{10}$, respectively.
To that order, however, nothing can be said about the
$\pi \pi$ scattering amplitude. The task of calculating the second
corrections has been taken up in Ref.\cite{bkmppn} $D_{1,2}$ admit an
expansion of the form
\begin{equation}
D_i = d_i^0 + d_i^1 \, \mu + d_i^2 \, \mu^2 + {\cal O}(\mu^3) \, \,
, \quad i=1,2 \, \, ,
\label{dexp}
\end{equation}
modulo logs and $\mu = M_\pi / m$. The upshot of the lengthy
calculation in Ref.\cite{bkmppn} is the following (the numbers given
here should be considered preliminary). The chiral
expansion for $D_1$ converges nicely and one thus is able to extract
the isospin two S--wave scattering length $a_0^2$,
\begin{equation}
a_0^2 = -0.052 \pm 0.013 \, \, ,
\label{a02}
\end{equation}
compatible with the one loop CHPT prediction of
$a_0^2 = -0.042 \pm 0.008$.\cite{gl84}
 In contrast, in case of
$D_2$ the terms of order $\mu^2$ are  large and have a  sizeable
uncertainty. Therefore, the small contribution from the $\pi \pi$
interaction can only be isolated with poor precision.
In particular, the excitation of the Roper resonance and its
subsequent decay into the nucleon and two pions in the S--wave
has to be understood much better and also certain LECs related to the
$\pi N$ scattering amplitude. The corresponding extracted value for
$a_0^0$ is
\begin{equation}
a_0^0 = 0.16 \pm 0.05 \, \, ,
\label{a00}
\end{equation}
which is compatible with the CA value of $0.16$ and the one--loop CHPT
prediction of $0.20 \pm 0.01$.\cite{gl84} As stated before, the
theoretical uncertainty is much smaller than the one deduced from the
$\pi \pi N$ threshold amplitude ${\cal A}_{10}$.
 These are good and
bad news. Of course, $a_0^0$ can be determined e.g. from $K_{\ell4}$
decays or pionic molecules,
so the reaction $\pi N \to \pi \pi N$ offers the complementary
information on $a_0^2$. A calculation of the next corrections to
$D_1$ would be very welcome to be able to further sharpen the
extraction of $a_0^2$. For all the details, see.\cite{bkmppn}

\section{Low-Energy Theorems in {\boldmath $\pi^0$} Photoproduction}

\noindent Space forbids to discuss in detail the interesting story of the
theoretical and experimental determinations of the S--wave multipole
$E_{0+}$ in neutral pion photoproduction off protons. Already in 1991 it was
shown\cite{bgkm} that in the derivation of the old "LET" some terms at
order $M_\pi^2$ had been overlooked. Unfortunately, the
reexamination of the
Saclay and Mainz data gave a value seemingly supporting this incomplete
expression. This has led many to try to resurrect or reinterpret the old
"LET". An educational discussion is given in Ref.\cite{cnpp}
 Well, there is yet another twist to the story. The new data from
SAL and MAMI are in the process of being analyzed and seem to indicate a
value of $E_{0+}$ much smaller (in magnitude) than
the old "LET" value and a less pronounced energy dependence in the first
15 MeV above threshold.\cite{jack}\cite{fuchs} If
confirmed, this would be a nice support of the CHPT calculations. Although the
expansion of the electric dipole amplitude in
powers of the pion mass is slowly
converging, the range of values predicted is definitively smaller
in magnitude than the presently believed empirical value of $(-2.0 \pm 0.2) \,
10^{-3} M_\pi^{-1}$. In fact, the preliminary analysis shown by Bernstein
at this symposium points toward a great {\it success} of CHPT -
only with the loop
contribution one is able to understand the result for $E_{0+}(\omega)$.

Almost as a by--product came the formulation of LETs for the slopes of
the P--wave multipole $P_{1,2}$ at threshold.\cite{bkmplet}
These are related to the more
commonly used electric and magnetic multipoles via
\begin{equation}
P_1 = 3 E_{1+} + M_{1+} - M_{1-} \, , \quad
P_2 = 3 E_{1+} - M_{1+} + M_{1-} \, \, .
\label{plet}
\end{equation}
The expansion of $P_{1,2}$ in powers of $M_\pi$ is quickly converging and thus
these quantities are a good testing ground for the chiral dynamics. The third
P--wave multipole $P_3 = 2M_{1+}+M_{1-}$ is dominated by the $\Delta (1232)$.
These statements are counterintuitive to many practitioniers in the field. It
is therefore important to stress that only with the new CV machines and
improved detector technology one can access this narrow window of chiral
physics. It is a sociologically interesting phenomenon how these facts are
ignored by some, who believe that nothing can't be learned any more from such
"old" physics. Quite the contrary is true. One can indirectly infer these
P--wave combinations from the unpolarized data and finds satisfactory agreement
with the predictions. However, some assumptions have to be made since the
system is underdetermined, so such results can only be considered indicative.
 A direct test of these P--wave LETs
will come once the reaction $\vec{\gamma} p \to \pi^0 p$ has been measured and
analyzed at MAMI. The extension to electroproduction can be found in
Ref.\cite{bkmplet2}

Finally, a remark on the resonance contributions is at order since that still
seems to be a red herring to some. In the $q^4$ calculation\cite{bkmplet} of
$\gamma p \to \pi^0 p$, three LECs appear,
one related to $P_3(\omega)$ and two
to $E_{0+}(\omega)$. The numerical value of the P--wave LEC can be quite
accurately understood from $\Delta(1232)$ and $\rho$--meson exchange. The old
Mainz data, however, led to a puzzle for the two S--wave LECs.\cite{bkmplet}
To the contrary, analyzing the new MAMI data,
one finds that the numerical values of
these two LECs obtained from a best fit to the total and differential cross
sections in the threshold region can be reproduced by resonance exchange to a
good accuracy. Thus, in the threshold region, resonances do {\it not} pose a
problem to CHPT, even not the close--by $\Delta$. This important lesson has
not yet penetrated the prejudices of most people.

\section{Outlook}

More accurate data are necessary to systematically explore the strictures of
the explicit and spontaneous chiral symmetry breaking in QCD. As has become
clear from the many interesting talks at this symposium, we will soon be in a
much better situation to pin down certain LECs and make more detailed
predictions. From the theoretical side, two major issues need clarification,
 these are the extension to the three--flavor case and the consistent
 implementation of isospin--breaking (via the quark mass
 differences and virtual photons).

\section{Acknowledgements}

I would like to thank the organizers for inviting me and their superbe work.
I am grateful to my collaborators V\'eronique Bernard and Norbert Kaiser for
sharing with me their insights into chiral dynamics and fruitful work together.

\end{document}